 \documentstyle[11pt]{article}
\def\ll{\label}
\def\re{\ref}
\def\c{\cite}

\def\r1{(\ref{$1})}

\def\ti{\tilde}

\def\ba{\begin{array}{c}}

\def\ea{\end{array}}

\def\ov{\over}
\def\ha{{1\over 2}}

\def\l{\left}
\def\l({\left(}
\def\r){\right)}
\def\r{\right}

\def\al{\alpha}

\def\be{\begin{equation}}
\def\bc{\begin{center}}
\def\ec{\end{center}}
\def\bit{\begin{itemize}}
\def\eit{\end{itemize}}
\def\ee{\end{equation}}
\def\ed{\end{document}}
\def\bea{\begin{eqnarray}}
\def\eea{\end{eqnarray}}
\def\efr{\end{flushright}}

\begin{document}
\title{New nonultralocal  quantum integrable  models
through gauge transformation} 

\author{
Anjan Kundu \footnote {email: anjan@tnp.saha.ernet.in} \\  
  Saha Institute of Nuclear Physics,  
 Theory Group \\
 1/AF Bidhan Nagar, Calcutta 700 064, India.
 }
\maketitle
\vskip 1 cm

\begin{abstract} 
One of the few schemes for  obtaining an integrable nonultralocal quantum  
model  is its possible generation from    an ultralocal   model 
 by a suitable gauge transformation. Applying this scheme we discover  two
 new nonultralocal models, which fit well into the
    braided Yang-Baxter relations ensuring their quantum integrability.  Our
first model is generated
  from a lattice Liouville-like system, while the  second one which is an
exact lattice version of the
  light-cone sine-Gordon  is gauge transformed from a model, which gives
also the  quantum  mKdV  for a different gauge choice.
\end{abstract}

\skip 0.5cm

 PACS numbers 02.30.Ik,
  02.20.Uw,
11.10.Lm
03.65.Fd

\vskip 0.8cm

\noindent 1. {\bf Introduction}

Extension from the classical to the quantum domain \cite{Faddeev} and exact
solution of a number of quantum models are undoubtedly a
 major achievement in the theory of
  integrable systems. However, truly speaking inspite of an 
 impressive list of such 
  systems \cite{kul-skly}, the success is limited   mostly to
 a  class of 
models  known as ultralocal (UL) models , for which 
  the representative Lax operators at different lattice
 points  must commute:
$ L^{ul}_{2k}(v)L^{ul}_{1j}(u) =L^{ul}_{1j}(u)L^{ul}_{2k}(v)
, \mbox {for } j \neq k.$ 
In formulating  their quantum integrability \c{Faddeev} this ultralocality
 plays a crucial  role, since  
 only under such constraint  the
quantum Yang-Baxter equation (YBE)
\be R_{12}(u-v)L^{ul}_{1j}(u)L^{ul}_{2j}(v)
=L^{ul}_{2j}(v)L^{ul}_{1j}(u)R_{12}(u-v), \quad
j=1, \ldots,N
\ll{rll} \ee
can be  lifted for  the monodromy matrix $T^{ul}_a(u) ={L}^{ul}_{aN}(u)\ldots
{L}^{ul}_{a1}(u) $, to its global form
\be R_{12}(u-v)T^{ul}_{1}(u)T^{ul}_{2}(v)
=T^{ul}_{2}(v)T^{ul}_{1}(u)R_{12}(u-v).
\ll{rtt} \ee
 Defining   $\tau(u)= tr(T(u))$ therefore the trace
identity:
$[\tau(u), \tau(v)]=0$ can be proved, which  is equivalent 
 to the 
integrability condition $[c_n, c_m]=0, $ for the set of conserved quantities
$c_n, n=1,2, \ldots $  .

On the other hand there
exists a rich class of    classical integrable 
 models, e.g.  mKdV, KdV, light-cone sine-Gordon,
 complex sine-Gordon,
derivative NLS, nonlinear $\sigma$ model etc. \cite{Maillet},
 which  violate  ultralocality condition  
and therefore make their  quantum description 
  through standard   YBE formulation difficult. 
Nevertheless starting  from eighties to early nineties 
  a number of nonultralocal (NUL) systems were discovered 
 showing quantum integrability \cite {korepin,Nijhof,
babelon,alekfad,reshet,goddard}.
  Thereafter except \c{Maillet91,hlav94} and few others there were not much 
serious attempts  for a considerable period of time
 to develop this important theory, until
 probably  \c{hlavkun96} where
 a braided 
YBE formulation was developed for NUL models 
and applied successfully to bring the 
  mKdV model  into this quantum integrable 
class \c{mkdv95}. The situation however   
became  more discouraging  in recent years with almost
no new discoveries of such models made 
  and no new theoretical achievements  taking place, 
 inspite of an urgent  need for such developments in this subject. 
 Only very recently   a mixed
left-right  component  quantum 
integrable  mKdV model  has been introduced with its possible connection
to perturbed CFT \c{mkdv02}.

In such a scenario therefore it is highly desirable to apply
some systematic scheme for  discovering   new 
quantum integrable  NUL 
 systems which can enrich this important  class
of   models.
  In fact 
{\it nonultralocalization} of  an ultralocal  model may serve
 as such a technique, where the quantum Lax operator 
of a NUL 
model can be constructed from that of an integrable  UL model  
 by a suitable operator dependent local  gauge transformation.  
This method, which is capable not only to  derive new models but also 
to reveal their direct relation with an UL model,
was  introduced  first possibly in \c{korepin}. In  subsequent works
 this scheme was implemented implicitly in
\c{alekfad},
  incorporated in a general
framework in  \c{hlavkun96} and  used explicitly in \c{gmkdv01}.
Our aim here is to apply the same scheme for generating   new 
integrable models, which  make  
 valuable 
   additions to the  existing
list of such   quantum integrable NUL  systems. In particular we discover
two models;
the first one  being  a new
discrete   quantum  NUL system  constructed 
by a simple gauge transformation
 from a lattice Liouville-like model.
  Our second model is
 an exact
lattice version of the  
  light-cone sine-Gordon (LCSG), which is 
  also new as a   quantum NUL model.
Interestingly  the LCSG is constructed through  gauge transformation from 
  the same UL model, which yields 
 the  quantum mKdV  \c{gmkdv01}, but for a different gauge choice.

We find that
the quantum  NUL models  we obtain 
  fit well into the  
   braided YBE formulation   \c{hlavkun96}
 and can be  solved exactly through  algebraic Bethe
 ansatz   modified  for the NUL models \c{mkdv95,mkdv02}.

\noindent 2. {\bf The scheme}: 

The idea of the  scheme is to start from the Lax operator $L^{ul}_j$ of an
 UL model and applying a  local 
 gauge transformation like \be D_{j+1}^{-1}L^{ul}_jD_{j}=L^{nul}_j
\ll{gt}\ee
  construct 
the Lax operator of a NUL model. However  to be 
a proper integrable  system the representative 
Lax operator of  the 
NUL model must satisfy the braided YBE \c{hlavkun96}
\begin{equation}
{R}_{12}(u-v)L^{nul}_{1j}(u)L^{nul}_{2j}(v)Z_{12}^{-1}
= L_{2j}^{nul}(v) L_{1j}^{nul}(u)Z_{21}^{-1}{R}_{12}(u-v).
\ll{bqybel}\end{equation}
 together with the nonultralocal  
braiding relation
\begin{equation}
 L^{nul}_{2 j+1}(v)Z_{12}^{-1}L^{nul}_{1 j}(u)
=L^{nul}_{1 j}(u) L^{nul}_{2 j+1}(v).
\ll{zlzl1u}\end{equation} 
It should be noted that compared to the general relations introduced in 
 \cite{hlavkun96} we have restricted only to the nearest neighbour braiding
 and
 taken  the equations  in a
  conjugate form to emphasize the
fact that  these  relations, though   look   differently 
are in fact equivalent and  lead to the same integrability for the NUL
systems. It is obvious that at the trivial limit of  
$Z=1$ the 
braided YBE  reduces to the standard  YBE   and the braiding 
relation  turns into the usual  ultralocality condition.

 The success of the gauge transformation
would  depend naturally  on the suitable choice of the gauge operators,
which must ensure  the above braided extensions.
 In fact  using the inverse of transformation (\re{gt}) directly
in the  YBE (\re{rll})   for
   $L^{ul}$, it can be shown through 
some simple algebraic manipulations that 
 in the simplest case the gauge operator $D_m$ that  transforms   
  (\re{rll}) into the braided YBE (\re{bqybel}) for
$L^{nul}$  must satisfy 
 the  conditions   \be
D_{aj}L^{nul}_{bj}D_{aj}^{-1}=L^{nul}_{bj}Z_{ab}^{-1},\ \ \ \ [R_{12},\ 
D_{1j}D_{2j}]=0. \ll{cond}\ee
Interestingly the ultralocality condition is   
transformed  automatically 
 to the braiding
 relation (\re {zlzl1u}), whenever the first of the 
 conditions (\re{cond}) holds.   
Note that 
 the corresponding monodromy
matrix  for the periodic discrete chain of size $N$
 would also be   gauge related through (\re{gt}) as  $
T^{nul}= {L}^{nul}_{N}(u)\ldots
{L}^{nul}_{1}(u)=
 D_{N+1}^{-1}T^{ul}D_{1}=D_{1}^{-1}T^{ul}D_{1}$ due to 
$D_{N+1}=D_{1}$. Therefore, 
        YBE 
(\re {rtt}) for $T^{ul}$  due to  
$D_{a1}T^{nul}_{bj}D_{a1}^{-1}=T^{nul}_{bj}Z_{ab}^{-1}, $
 induced 
by  
(\re{cond}),  yields the  global braided  YBE
\begin{equation}
{R}_{12}(u-v)T^{nul}_{1}(u)Z_{21}^{-1}T^{nul}_{2}(v)Z_{12}^{-1}
= T_{2}^{nul}(v)Z_{12}^{-1} T_{1}^{nul}(u)Z_{21}^{-1}{R}_{12}(u-v).
\ll{bqybet}\end{equation}
associated with the NUL models. In \cite{hlavkun96} 
the structure  of the braiding matrices $Z$ for which 
 (\re{bqybet}) leads to the integrability condition has been analyzed.
In the present applications  however we show  the integrability 
by directly  using  in 
(\re{bqybet})  the explicit
 form of the braiding
matrices,  which are  much  simpler in our case.   

Since the NUL model 
must share the same $R$-matrix with the UL model, from which it is to be 
  generated, we  search 
for such a suitable 
source model associated with the well known $4\times 4$ trigonometric $R$-matrix,
 the nontrivial 
  elements of which are
\begin {equation} 
R^{11}_{11} = R^{22}_{22}= \sin \al (u+1),
\  R^{12}_{12} = R^{21}_{21}= \sin \al u, \ R^{12}_{21} = R^{21}_{12}=
 \sin \al . 
        \ll{R-mat}\end {equation}
The  general discrete  Lax operator of the UL systems related to (\re{R-mat})  
 as shown in \c{kunprl99}  may be given by 
\be
L_k{(u)} = \left( \begin{array}{c}
  {c_+^+}{\xi} q^{ S_k^3}+ {c_+^-} {1 \ov \xi} q^{- S_k^3}\qquad \ \ 
  2\sin \al  S_k^-   \\
    \quad  
   2\sin \al S^+_k    \qquad \ \  {c_-^+}{\xi} q^{- S^3_k}+ 
{c_-^-}{1 \ov \xi} q^{S^3_k}
          \end{array}   \right), \quad
         q=
e^{i \alpha } , \ \ {\xi}= e^{i \alpha u } \ll{L} \ee
 linked with
 the underlying generalized quantum
algebra
\be
 [S_k^3,S_l^{\pm}] = \pm \delta_{kl} S_k^{\pm} ,  \ \ [ S_k^ {+}, S_l^{-} ] =
  \delta_{kl}( M^+[2  S^3]_q + {M^- } 
[[ 2  S_k^3 ] ]_q) , \quad  [M^\pm, \cdot]=0.
\ll{Alg} \ee
 Here $ [x]_q \equiv
{\sin (\al x)\over \sin \al},\ \ [[x]]_q \equiv
{\cos (\al x)\over \sin \al}
$ and the central elements
 $ M^\pm$ are related to the other set of  such elements appearing in the 
$L$-operator 
as
  $ M^\pm=\pm \ha  \sqrt {\pm 1} ( c^+_+c^-_- \pm
c^-_+c^+_- ). $  The ultralocality condition for (\re{L}) 
holds  naturally  due to the algebraic relations 
(\re{Alg}), which is  valid only at the same lattice sites.
It is important to note that since  condition (\re{cond}) is a linear
relation for the Lax operator, the same condition must  hold also 
for the UL source model, provided the braiding matrix $Z$ commutes
with the gauge operator $D_j.$  
Therefore our strategy is to seek for the suitable   structure of an
 UL Lax operator as some realization of (\re{L}), such that it also satisfies
 (\re{cond}),   crucial   for generating NUL models.
We are able to find 
 two such  structures  for  two different sets of  choice 
 of the    central
elements in (\re{L}).
\\
\noindent 3. {\bf NUL model  from the Liouville model}:

 Lattice Liouville model (LLM)
 \cite{llm1} can be derived from the general Lax operator 
(\re{L}) for the choice $c^-_+= c^+_-=0,\ \  c^+_+= c^-_-=1,$ which reduces 
the second algebraic relation of (\re{Alg})  to 
\be
 [ S_k^ {+}, S_l^{-} ] =
  -2i\delta_{kl} \ e^ {2 i\al S_k^3}   \sin \al
\ll{eAlg} \ee
and allows  a realization of the generators as
\be
 S_k^ {+}=e^{-ip_k}g(u_k), \ S_k^ {-}=g(u_k)e^{ip_k}, \ S_k^3=u_k, \
\ \ g(u)=(\kappa+ \Delta e^{i\al (u+\ha)})^\ha \ll{llm}\ee
  in canonical variables  $[u_k,p_l]= i \delta_{kl}.$ Note that for any
arbitrary value of $\kappa$  and under any canonical transformation
  (\re{llm}) remains a proper realization of (\re{eAlg}) and therefore
 retains the integrability of
the model. 
The well known LLM is obtained   at
 $\kappa=1,$ while 
for the generation of our new  model  we choose $\kappa=0$
 along with a trivial  canonical transformation 
$p \to \al u, \ \al u  \to -p .$ This gives the Lax operator of our
LLM
  as a realization of (\re{L})  in the form
\be
L_k{(u)} = \left( \begin{array}{c}
  \xi e^{-ip_k}\qquad \ \ 
   e^{i(\al u_k-p_k)}  \\
    \quad  
    e^{-i(\al u_k+p_k)} \qquad \ \   
{1 \ov \xi} e^{-ip_k}
          \end{array}   \right), \quad
          \ll{Lul1} \ee
which would serve now as our 
 source 
UL model.
 Since  NUL  Lax operators 
 usually involve derivatives of the canonical variables or 
 operators with  current-like commutation relations
$ [ v(x),v(y) ]=\pm i \al \delta'(x-y),$ we introduce  for our 
lattice  construction  the
discrete version of such fields $ v^\pm_k$,  defining  their
  commutators in the form 
\be
[v^\pm_k, v^\pm_l]= \pm i {\al \ov 2} (\delta_{k+1,l}- \delta_{k,l+1}).
\ll{cr1}\ee
 The fact that these {\it nonultralocal } fields can
 be realized through canonical variables as
\be v^+_k=-{\al \ov 2}(u_{k+1}-u_{k}) -p_k, \ \ 
v^-_k={\al \ov 2}(u_{k+1}-u_{k}) -p_k 
\ll{map}\ee
would play a significant role in our construction.

Remarkably, the structure of (\re{Lul1}), more
specifically  the 
same exponential dependence   as $\sim e^{-p_k}$ for all its elements
permits us to choose a simple gauge operator in the form $ D_j=
e^{-i {\al \ov 2} u_j \sigma ^3}$. Applying    this matrix gauge operator
in  transformation
(\re{gt}) on the  LLM-like  Lax operator 
(\re{Lul1})   and
 changing from canonical to
current-like  variables  as in  (\re{map})  we can
 generate finally    the  Lax operator of our   quantum
integrable
 NUL model.   
  
Straightforward calculations yield  the explicit form of
 this $L$-operator 
  as
\be
L^{(1)}_k{(u)} = \left( \begin{array}{c}
  \xi \  W_k^+\qquad \ \ 
   W^+_k  \\
    \quad  
    W^-_k \qquad \ \   
{1 \ov \xi} \ W^-_k
          \end{array}   \right), \quad
W^\pm_k=e^{i v^\pm_k}, \ \ \xi =e^{i \al u}          , \ll{Lnul1} \ee
expressed completely through  
 variables $ v^\pm_k$ having  {\it nonultralocal} commutation
 relations (\re{cr1}).

 With all  required objects in hand  we must check now
 that  the essential conditions (\re{cond})
are    fulfilled in our construction.
Indeed we find that  the first  of these  conditions is 
satisfied by both
the Lax operators (\re{Lul1}) and (\re{Lnul1})
  yielding the explicit form of the braiding matrix $Z$ as
\be Z_{12}=e^{i{ \al \ov 2} \sigma^3_1}. \ll{Z1}\ee
The second  condition on the other hand  holds due to the symmetry of the 
 trigonometric $R$-matrix  (\re{R-mat}) and the specific form of 
our gauge operator.
Consequently, as shown
above,    Lax operator (\re{Lnul1}) should also satisfy the
braided YBE relations (\re{bqybel}), (\re{zlzl1u}) and (\re{bqybet}).
The associated $R$-matrix is given by the same 
(\re{R-mat}) and the braiding matrix as in (\re{Z1}). 
 Thus we conclude that the Lax operator (\re{Lnul1}) we have generated 
from a LLM-like model (\re{Lul1})
 represents a new  integrable  quantum 
NUL  model. The full set
 of conserved quantities for this integrable  periodic  model 
 may be obtained directly
from   (\re{bqybet})   through
diagonal elements $A(u), D(u) $ of $
T^{nul}(u)$ as $\ti \tau(u) =q^{-\ha}A(u)+q^{\ha}D(u)= \sum_n^N
 \xi^{\pm n}
c_{\pm n}, \ \ \ \xi =e^{i \al u} $. We may
 define the Hamiltonian of the model as $H= c_{N-2}(c_N)^{-1} +
c_{-(N-2)}(c_-N)^{-1}
= 2\sum_j \cos (v^-_j -v^+_j -{\al \ov 2})$. For
finding the dynamical equations  however we have to use the 
noncanonical commutation relations (\re{cr1}) together with 
\be
[v^+_k, v^-_l]=i {\al \ov 2} (\delta_{k+1,l}- 2\delta_{k,l}+\delta_{k,l+1}),
\ll{cr2}\ee
which is also compatible with the realization (\re{map}). Note the
intriguing fact that the noncommutativity of the operators (\re{cr1}) and
(\re{cr2}) at different
sites induce   nearest-neighbour interaction in the model.   \\ 
\noindent 4. {\bf Light-cone sine-Gordon as NUL 
 model}:

For constructing  our next model we choose
$c^-_+= c^-_-=0,\  c^+_+= c^+_-= \Delta$ or its complementary set
$c^-_+= c^-_-=\Delta, \ c^+_+= c^+_-= 0,$ where $\Delta$ is the lattice
constant. Evidently  both of 
 these choices give $M^\pm=0$ and reduce 
 (\re{Alg})  to a simplified algebra
\be
 [ S_k^ {+}, S_l^{-} ] =0, \ 
\  [S_k^3,S_l^{\pm}] = \pm \delta_{kl} S_k^{\pm}, 
\ll{eAalg2} \ee
   which may be realized as 
$ S_k^ {+}=(S_k^ {-})^{-1}=e^{-ip_k},  \ \al S_k^3=\al u_k \mp p_k .\ $
The Lax operators with the above  two sets of choices for the  central
elements and using (\re{eAalg2}) therefore  can be realized 
    from   (\re{L}) 
in the form
\be
L^{(-)}_k{(u)} = \left( \begin{array}{c}
  \Delta \xi e^{i(\al u_k-p_k)}\qquad \ \ 
   e^{ip_k}  \\
    \quad  
    e^{-ip_k} \qquad \ \   
\Delta \xi e^{-i(\al u_k-p_k)}
          \end{array}   \right), \quad 
L^{(+)}_k{(u)} = \left( \begin{array}{c}
  \Delta {1 \ov \xi} e^{-i(\al u_k+p_k)}\qquad \ \ 
   e^{ip_k}  \\
    \quad  
    e^{-ip_k} \qquad \ \   
\Delta {1 \ov \xi} e^{i(\al u_k+p_k)}
          \end{array}   \right),
          \ll{Lul2-+} \ee
  representing two quantum integrable UL lattice models,
which we intend to
use as our source model for constructing  NUL systems.
 Note that the {\it right} operator 
$L^{(+)}_k{(u)}$ can be obtained formally from the {\it left} one 
$L^{(-)}_k{(u)}$  through a simple mapping ${ \xi} \to {1 \ov \xi},
\al \to - \al $. Therefore we  deal explicitly with the 
{\it left} case only and recover the results related to the complementary
{\it right} case through the above  mapping.
Interestingly, one can choose the 
 gauge operator $\ti  D_j $ in the present  case   as the {\it square}
 of that
  considered above for the LLM, i.e. $ \tilde  D_j (\al)= D^2_j=
e^{-i {\al } u_j \sigma ^3}\ \ $ for the {\it left} model and 
applying  gauge transformation (\re{gt}) as $q^{-\ha}\tilde  D_{j+1} (\al)
 \ (L^{(-)}_j{(u)}\sigma^1) \ 
\tilde  D^{-1}_j(\al)=L^{(-)lcsg}_j{(u)}$
 construct the  Lax operator 
\be
L^{(-)lcsg}_j{(u)} = \left( \begin{array}{c}
   e^{i(p_j-\al \nabla u_j)}\qquad \ \ 
  \Delta \xi e^{-i(p_j+\al  u_{j+1})}  \\
    \quad  
    \Delta \xi e^{i(p_j+\al  u_{j+1})} \qquad \ \   
e^{-i(p_j-\al \nabla u_j)}
          \end{array}   \right), \quad \nabla u_j \equiv  u_{j+1}- u_{j} 
\ll{lcsg}\ee
with $\xi =e^{i \al u}.$
Note that (\re{lcsg})  is the Lax operator of 
 an exact lattice version of the LCSG model and
represents  indeed a new  quantum NUL model.
We can check directly 
   that both  the source UL model
$L^{(-)}_k{(u)}$ in (\re {Lul2-+}) and the $L$-operator of the  
 NUL model (\re{lcsg})
respect  conditions (\re{cond})  required for the scheme,
giving the  braiding
matrix in the explicit form
\be Z_{12}^{(-)lsgg}=e^{i \al \sigma^3 \otimes \sigma^3}. \ll{Z2}\ee Therefore 
the Lax operator (\re{lcsg})  satisfies also the braided YBE
 (\re{bqybel}),  (\re{bqybet})   and the  nonultralocal relation
(\re{zlzl1u}) with   $R$-matrix (\re{R-mat}) and  $Z$-matrix  (\re{Z2}),
 which
 establishes the quantum integrability of the NUL model
 we have generated. 
The significance of the model is revealed 
at the continuum limit ${\Delta \to 0} $, when  it reduces 
to  the light-cone sine-Gordon 
 field model. At this field limit 
  we have $p_j \to \Delta \partial_t
u(x), \
  \al u_{j} \to u(x),\ \al u_{j+1} \to u(x) +\Delta  \partial_x
u(x) $. Defining therefore $ \partial_t  
u \pm \partial_x
u =\ha  \partial_\pm 
u$ in the light-cone coordinates it is not difficult to show that 
the lattice LCSG (\re{lcsg}) reduces to the well known  field operator form
 $L^{(-)lcsg}_j{(u)} \rightarrow  
  I +\Delta \sc {U}_-(x)$, where $ \sc {U}_-(x) 
=\left( \begin{array}{c}
i \ha \partial_- u(x)   \qquad \ \ 
   \xi e^{-i  u (x)}  \\
    \quad  
    \xi e^{i  u (x)} \qquad \ \   
-i \ha \partial_- u(x)
          \end{array}   \right),
$ yields  one of the well known 
 Lax pair: $\partial_- \Phi=\sc {U}_- \Phi$.

It is intriguing to follow  that a complementary {\it right} LCSG 
 similarly may be generated from  $L^{(+)}_k{(u)} $ in  (\re{Lul2-+})
as 
$q^{\ha}\tilde  D_{j+1} (-\al)
 \ (L^{(+)}_j{(u)}\sigma^1) \ 
\tilde  D^{-1}_j(-\al)=L^{(+)lcsg}_j{(u)}$, which represents  again  a new
 quantum
integrable NUL model associated with the same R-matrix, while 
 the braiding matrix is  
given by $Z_{12}^{(+)lsgg}=e^{-i \al \sigma^3 \otimes \sigma^3}.$
 Note that the NUL model  $L^{(+)lcsg}_j{(u)} $ can  be obtained formally from 
$L^{(-)lcsg}_j{(u)}$  by using  the same mapping 
${ \xi} \to {1 \ov \xi},
\al \to - \al $  and  
yields
 at the continuum
 limit $L^{(+)lcsg}_j{(u)} \rightarrow  
  I -\Delta \sc {U}_+(x)$, with  $ \sc {U}_+(x) $ being the other component
of the Lax pair. Zero curvature condition: \  $ 
\partial_- {U}_+ -\partial_+ {U}_+ + [ {U}_+, {U}_-]=0$ 
 involving both Lax operators gives  the well known form of the sine-Gordon field 
equation in light-cone coordinates: $ \partial^2_{+-}u=2\sin2 u$. 

As it has been shown recently \c{gmkdv01}, the NUL quantum mKdV$^{(\pm)}$ models can also
be obtained by gauge transforming some UL models . It is
remarkable  that these UL models may be given exactly by
the same source models $L^{(\pm)}_k{(u)} $   (\re{Lul2-+}) found here 
  for the LCSG. However the
gauge operator required for constructing the  mKdV
should be given  by $   D_j (\al)=(\tilde D_j(\al))^\ha=
e^{-i \ha{\al } u_j \sigma ^3}$ and
 therefore the braiding matrix for the 
mKdV is related similarly to that of the LCSG as $Z^{(\pm)mkdv}
=(Z^{(\pm)lcsg})^{\ha}
 =q^{\ha \sigma^3 \otimes \sigma^3}$, as can be confirmed by
comparing \c{gmkdv01} with our result.
It is intriguing to note that, the gauge operator $   D_j (\al) $ for
the mKdV model coincides on the other hand    
 with that  
for our new nonultralocal LLM-type model, we have found above.

 For exact solution of 
 the eigenvalue problem of the  Hamiltonian and
 higher conserved operators for  quantum NUL systems, one needs  
   modification 
of the algebraic Bethe ansatz (ABA). Such a   
formulation for  the NUL mKdV has already been developed in
\c{mkdv95,mkdv02}.
Since the quantum NUL systems that 
we have discovered here are very close in structure to the quantum mKdV,
 we can apply successfully the  modified ABA to the present models
 following closely the steps of the quantum mKdV and therefore we
 omit them here. 
 
\noindent 5. {\bf Conclusion}

It seems that the gauge generation scheme adopted here for constructing 
new quantum integrable  NUL models starting from ultralocal ones 
is far more promising than might have been expected and applied for, so far.
 Our success encourages us
to look for \c{kun02}  the explicit 
applicability of this scheme   to the well known NUL models, e.g.
 quantum mapping \c{Nijhof}, Coulomb-gas CFT related models \c{babelon},
 WZWN model \c{alekfad} etc.
It  would certainly be desirable to exploit this scheme 
    for solving the challenging   problem of establishing  the 
braided YBE formulation for NUL models like nonlinear $\sigma$-model,
derivative
NLS, complex sine-Gordon model etc.


\begin{thebibliography}{99}
\bibitem{Faddeev} L. D. Faddeev, Sov. Sc. Rev. C1 (1980) 107.
\bibitem{kul-skly} P. Kulish and E. K. Sklyanin,
Lect. Notes in Phys. (ed. J. Hietarinta et al, Springer,Berlin, 1982) vol. 151
p. 61.
\bibitem{Maillet} 
S. A. Tsyplyaev, Teor. Mat. Fiz. 46 (1981) 24

J.M. Maillet,  Phys. Lett. 162 B (1985) 137;
;
 Nucl. Phys. B269 (1986) 54 ; Phys. Lett. 167 B (1986) 401.

M. Semenov-Tian-Shansky, Funct. Anal. Appl. 17 (1983) 259

\bibitem{korepin} V. E. Korepin, J. Sov. Math. 23 (1983) 2429.
 \bibitem{Nijhof} F.W. Nijhoff, H.W. Capel and V.G. Papageorgiou,
{\it Integrable quantum mappings },
Clarkson Univ. preprint INS 168/91 (1991)

 F.W. Nijhoff and H.W. Capel,
 {\it Integrable quantum mapping and nonultralocal Yang-Baxter structure},
Clarkson Univ. preprint INS 171/91 (1991).
\bibitem{babelon} O. Babelon and L. Bonora, Phys. Lett. 253 B (1991) 365
;
 O. Babelon, Comm. Math. Phys.  139 (1991) 619
;
 L. Bonora and V. Bonservizi, Nucl. Phys. B 390  (1993) 205.

\bibitem{alekfad} A. Alekseev, L.D. Faddeev, M. Semenov-Tian-Shansky 
 Comm. Math. Phys.  149 (1992) 335
;
L. D. Faddeev,  Comm. Math. Phys.  132 (1990) 131
;
A. Alekseev, S. Statashvili, Comm. Math. Phys.  133 (1990) 353
;
B. Blok,  Phys. Lett. 233B (1989) 359

\bibitem{reshet} N. Yu. Reshetekhin and  M. Semenov-Tian-Shansky ,
 Lett. Math. Phys. 19  (1990) 133
\bibitem{goddard} M. Chu, P. Goddard, I. Halliday, D. Olive and
A. Schwimmer,
 Phys. Lett. 266 B  (1991) 71.

\bibitem{Maillet91} L. Freidel and J.M. Maillet, Phys. Lett. 262 B  (1991) 278
;  263 B  (1991) 403
\bibitem{hlav94} L. Hlavaty,  J. Math. Phys. 35 (1994) 2560
\bibitem{hlavkun96}
L. Hlavaty and Anjan Kundu Int. J. Mod. Phys. A11 (1996) 2143
\bibitem{mkdv95}
Anjan Kundu, Mod. Phys. Lett. {A10} (1995) 2955,
\bibitem{mkdv02}
D. Fioravanti and M. Rossi, J. Phys. A 35 (2002) 3649
\bibitem{gmkdv01} D. Fioravanti and M. Rossi, J. Phys. A 34 (2001) L567 
\bibitem{kunprl99}
   Anjan Kundu, Phys. Rev. Lett. 82   (1999) 3936
\bibitem{llm1}
L.D. Faddeev, O. Tirkkonen, Nucl. Phys. {B453} (1995) 647
\bibitem{kun02} Anjan Kundu, {\it under preperation} 

\end{thebibliography}
\end{document}